\documentclass[12pt,thmsa]{article}
\usepackage{amssymb}
\usepackage{amsmath}
\usepackage{amsfonts}
\usepackage{graphicx}

\newcommand{\bra}[1]{\bigl\langle #1 \bigr|}
\newcommand{\ket}[1]{\bigl| #1 \bigr\rangle}
\newcommand{\expect}[1]{\left\langle #1 \right\rangle}
\makeatletter

\begin{document}

\begin{center}
{\Large Fisher information of two superconducting charge qubits under dephasing noisy} \\[0pt]
Hamada El-Shiekh$^{a,b}$, Nour Zidan$^{a}$ and Nasser
Metwally$^{c,d}$\\
 $^{a}${\small Mathematics Department, Faculty of
Science, Sohag University, Sohag, Egypt.} \\$^{b}${\small
Mathematics Department,
College of Science, Al-Taif University, Saudi Arabia.}\\
$^c${ Department of Mathematics, College of Science, Bahrain
University, Bahrain. }\\
$^d${Department of Mathematics, Faculty of Science, Aswan
University, Aswan, Egypt.}
\end{center}
\begin{abstract}

The dynamics of  a charged two-qubit system prepared  initially in
a maximum entangled  state is discussed, where  each qubit
interacts independently with a dephasing channel. The Fisher
information is used to estimate the channel and the energy
parameters. Moreover, the contribution from different parts of Fisher
information is discussed. We show that,  the  degree of estimation
of any parameter depends on the initial value of this parameter.
It turns  out that the upper bounds of Fisher information with respect to the
dephasing parameter increase and that  corresponding to coupling
parameter decreases as the initial energy of each qubit increases.

\end{abstract}
\section{Introduction}

\bigskip Quantum Fisher information  is considered  as one of the important
measurements in the estimation theory \cite{in1,in2}, which
characterizes the sensitivity of a given system with respect to
the changes in one of its parameters. Recently, in the context of
quantum estimation theory Fisher information has paid  some
attentions. For example, Z. Qiang et. al \cite{in3} examined the
problem of parameter estimation for two initially entangled qubits
subject to decoherence . Xiao et. al.\cite{in4} showed that the
partial measurements can greatly enhance the quantum Fisher
information teleportation under decoherence. \ F. Fr\"{o}wis et.
al  proposed an experimentally accessible scheme to determine
lower bounds on the quantum Fisher information\ \cite{in5}.
Quantum Fisher information of the Greenberger-Horne-Zeilinger
state in decoherence channels was discussed by Ma et.al
\cite{in6}. The dynamic of quantum Fisher information of W state
in the three basic decoherence channels was studied by Ozaydin
\cite{in7}. The properties of quantum Fisher information in a
general superposition of a 3-qubit GHZ state and two W states were
investigated by Yi et. al \cite{in9}. The time evolution of the
quantum Fisher information of a system whose the dynamics was
described by the phase-damped model is discussed by Obada et. al
\cite{in10}. The effect of the Unruh parameter on the precision of
estimating the channel parameter was discussed in
\cite{Metwally2016-1}.  Recently, Metwally\cite{Metwally2016-2}
 employed the Fisher information to estimate the teleported and
the gained parameters by using an accelerated channel.

   However, investigating the Fisher information for the charged qubits  has not paid  much attention.
    Therefore, we are motivated to
   use the Fisher information to  estimate the initial parameters  of the
   charged qubits system; the Josephon energies parameters and the mutual coupling energy between the two
   qubits and the channel noise parameter.

 This paper is organized as follows: In Sec. 2, we present the
model and its solution under dephasing noisy.  In Sec. 3, we
estimate  the charged system's parameters and  the noisy channel
parameter by means of Fisher information, where analytical forms
are introduced. We summarize our results and conclusion in Sec. 4.

\section{The model and its solution}

Charged qubit represents one of the most promising particles in
quantum computation and information. There are several studies
have been introduced to investigate the properties of entangled
charged qubits.  For example, Liao et. al, \cite{in12}
 investigated the entanglement between two Josephson charge
 qubits. The dynamics of charge qubits coupled to a nonmechanical
 resonator under the influence a phonon bath is discussed by Abdel-Aty et.al
 \cite{in15}. Dynamics of the quantum deficit of two charged qubit
 is investigated in \cite{Metwally2009}. The possibility of using
 charged qubit to perform quantum teleportation is investigated by
 Metwally \cite{Metwally2009-1}.

 However, the  Hamiltonian which can be
used to describe a two charged qubit system is defined as
\cite{m1,m2,m3}:
\begin{equation}
\mathcal{H}=-\frac{1}{2}\Bigl\{\kappa_1\sigma_z^{(1)}\otimes
I_2+\kappa_2I_1\otimes\sigma_z^{(2)}+E_{J_1}\sigma_x^{(1)}\otimes I_2
+E_{J_2}I_1\otimes
\sigma_x^{(2)}-2E_m\sigma_z^{(1)}\otimes\sigma_z^{(2)}\Bigr\}
\end{equation}
where
$\kappa_1=2E_{c_1}(1-2n_{g_1})+E_m(1-2n_{g_2})$,~$\kappa_2=2E_{c_2}(1-2n_{g_2})+E_m(1-2n_{g_1})$,
$ I_1,I_2$ are the unit operators for the first and the second
qubit, respectively and  $E_{c_i},~E_{J_i}$ are the charging and
Josephon energies for the first and the second qubit,
respectively. The mutual coupling energy between the two qubits is
defined by the parameter $E_m$. The operators
$\sigma_{z,x}^{(i)},~i=1,2$ represent the Pauli operators  for the
two qubits, respectively.

 Assume that the charged qubit state is
prepared in the maximum Bell state
$\rho_{\psi^+}=\ket{\psi^+}\bra{\psi^+}$,
$\ket{\psi^{+}}=\frac{1}{\sqrt{2}}(\ket{01}+\ket{10})$. In our
investigation, we consider the following assumptions:
\begin{enumerate}
\item[(i)] The two charged qubits are identical, namely, $%
E_{J_{1}}=E_{J_{2}}=E_{J}$.

\item[(ii)]The system is assumed to be is in the degenerate point,
namely,  $n_{g_{1}}=n_{g_{2}}=0.5$.

\item[(iii)] The noise channels are identical, i.e., having  the
same strength.
\end{enumerate}
 Under these considerations, the Hamiltonian  (1)  reduces
to be,
\begin{equation}
\mathcal{H}=-\frac{1}{2}\Bigl\{E_{J_1}\sigma_x^{(1)}\otimes I_2
+E_{J_2}I_1\otimes
\sigma_x^{(2)}-2E_m\sigma_z^{(1)}\otimes\sigma_z^{(2)}\Bigr\}.
\end{equation}
 The master equation which governed the system is given by:

\begin{equation}
\frac{d\rho (t)}{dt}=-i\left[ \mathcal{H},\rho \right]
+\frac{\Gamma}{8} \sum_{j=1,2}\left( 2\sigma_{z}^{(j)}\rho
\sigma_{z}^{(j)}-\sigma_{z}^{(j)}\sigma_{z}^{(j)}\rho -\rho
\sigma_{z}^{(j)}\sigma_z^{(j)}\right) \label{m3}
\end{equation}
where, $\Gamma$ is the dephasing parameter.  By solving this
master equation, one obtains  $\rho(t)$, which describes the
evolution of the charged system. This state is described by a
square matrix of order 4 with elements are given by:
\begin{eqnarray}
\varrho_{11}&=&\varrho_{44}=\Upsilon_1\Bigl\{2\lambda_{123}\exp[2\Gamma
t]-\lambda_2\mu_{-}R^{+}_1-\lambda_1\mu_{+}R_2^{+}\Bigr\},
 \nonumber\\
\varrho_{22}&=&\varrho_{33}=\Upsilon_1\Bigl\{2\lambda_{123}\exp[2\Gamma
t]+\lambda_2\mu_{-}R^{+}_1+\lambda_1\mu_{+}R^{+}_2\Big\},
\nonumber\\
\varrho_{14}&=&\varrho_{41}=\Upsilon_1\Bigl\{2\lambda_{123}\exp[-2\Gamma
t]+\lambda_2\mu_{-}R^{-}_1+\lambda_1\mu_{+}R^{-}_2\Bigr\},
\nonumber\\
\varrho_{23}&=&\varrho_{32}=\Upsilon_1\Bigl\{2\lambda_{123}\exp[-2\Gamma
t]-\lambda_2\mu_{-}R^{-}_1-\lambda_1\mu_{+}R^{-}_2\Bigr\},
\nonumber\\
\varrho_{12}&=&\varrho_{13}=\varrho_{42}=\varrho_{43}=\Upsilon_2\Bigl\{2E_{m} \left[ \cosh \left(
\sqrt{2}\lambda _{2}t\right) -\cos\left( \sqrt{2}\lambda
_{1}t\right) \right]
\nonumber\\
&&+i\sqrt{2}\left[\lambda _{2}\sinh \left( \sqrt{2}\lambda
_{2}t\right) +\lambda _{1}\sin \left( \sqrt{2}\lambda _{1}t\right)
\right] \Bigr\},\nonumber\\ \varrho_{21}&=&\varrho_{31}=\varrho_{24}=\varrho_{34}=\varrho^*_{12}
\end{eqnarray}
where,
\begin{eqnarray}
 \Upsilon_1&=&\exp(-2\Gamma t)/8\lambda_{123},\quad\Upsilon_2= \frac{E_{j}\exp (-2\Gamma t)}{8\lambda
 _{3}},\quad
 \lambda_{123}=\lambda_1\lambda_2\lambda_3,\quad
\nonumber\\
 \mu_{\pm}&=&\lambda_3\pm(\Gamma^2+E_m^2-E_J^2)
\nonumber\\
R^{\pm}_1&=&\sqrt{2}\Gamma\sin(\sqrt{2}\lambda_1t)\pm\lambda_2\cos(\sqrt{2}\lambda_1t),
\nonumber\\
R^{\pm}_2&=&\sqrt{2}\Gamma\sinh(\sqrt{2}\lambda_2t)\pm\lambda_2\cosh(\sqrt{2}\lambda_2t),
\nonumber\\
\lambda _{1,2} &=&\sqrt{\lambda _{3}\pm\left(
E_{m}^{2}+E_{j}^{2}-\Gamma ^{2}\right) },\quad \nonumber\\
 \lambda
_{3}&=&\sqrt{E_{m}^{4}+\left( E_{j}^{2}-\Gamma ^{2}\right)
^{2}+2E_{m}\left( E_{j}^{2}+\Gamma ^{2}\right) ^{2}}.
\end{eqnarray}
The eigenvalues of the density operator (4) are given by:
\begin{equation}
\epsilon _{1}=\varrho _{11}-\varrho _{14},\text{ \ }\epsilon _{2}=\varrho
_{22}-\varrho _{23},\text{ \ \ }\epsilon _{3,4}=-2\sqrt{\alpha ^{2}+\beta
^{2}}\mu _{\pm }
\end{equation}
 and the corresponding  eigenvectors
are given by:
\begin{eqnarray}
\left\vert V_{1}\right\rangle &=&\frac{1}{\sqrt{2}}\left( -1,0,0,1\right),%
\text{ \ \ \ }\left\vert V_{2}\right\rangle =\frac{1}{\sqrt{2}}\left(
0,-1,1,0\right) ,
\nonumber\\
\left\vert V_{3}\right\rangle &=&\frac{1}{\sqrt{2(1+\mu_{-}
^{2})}}\left( 1,~\mu_{-} \exp (-i\theta ),~\mu_{-} \exp (-i\theta
),1\right) ,\text{ \ \ \ \ \ \ \ } \nonumber\\
\left\vert
V_{4}\right\rangle &=&\frac{1}{\sqrt{2(1+\mu_{+} ^{2})}}\left(
1,~\mu_{+} \exp (-i\theta ),~\mu_{+} \exp (-i\theta ),1\right),
\end{eqnarray}
where,

\begin{eqnarray}
\theta &=&\tan ^{-1}\left( \frac{\beta }{\alpha }\right) ,
\nonumber\\
\alpha &=&\frac{E_{j}E_{m}\exp (-2\Gamma t)}{4\lambda
_{3}}\left[ \cosh
\left( \sqrt{2}\lambda _{2}t\right) -\cos\left( \sqrt{2}\lambda _{1}t\right) %
\right] ,  \notag \\
\beta &=&\frac{\sqrt{2}E_{j}\exp (-2\Gamma t)}{8\lambda _{3}}\left[ \lambda
_{2}\sinh \left( \sqrt{2}\lambda _{2}t\right) +\lambda _{1}\sin \left( \sqrt{%
2}\lambda _{1}t\right) \right] \nonumber\\
\mu_{\pm} &=&\frac{1}{4\sqrt{\alpha ^{2}+\beta ^{2}}}\left( -\varrho
_{11}-\varrho _{14}+\varrho _{22}+\varrho _{23}\pm\sqrt{(\varrho _{11}+\varrho
_{14}+\varrho _{22}+\varrho _{23})^{2}+16\varrho _{12}\varrho _{21}}\right).
  \nonumber\\
\end{eqnarray}

\section{Fisher information}
In the estimation theory, the quantum Fisher information plays a
central role. There are some parameters cannot be quantified
directly, so quantum
Fisher information can be used to estimate these parameters \cite{m6}. Let $%
\eta $ represents the parameter to be estimated and the density operator of
the spectral decomposition is given as $\varrho _{\eta }=\sum_{i=1}^{n}\mathcal{%
E}_{i}\left\vert \mathcal{V}_{i}\right\rangle \left\langle \mathcal{V}%
_{i}\right\vert $  where, $\mathcal{E}_{i}$ and $\left\vert \mathcal{V}%
_{i}\right\rangle $ are the eigenvalues and eigenvectors of of the
density operator $\rho _{\eta }$, respectively. The Fisher
information corresponding to the  parameter $\eta$  is given by
(see \cite{in1,Metwally2016-1,m9,m10}):
\begin{equation}
\mathcal{F}_{\eta}=\mathcal{F}_{C}^{\eta}+\mathcal{F}_{P}^{\eta}-\mathcal{F}_{M}^{\eta},  \label{q2}
\end{equation}
where,
\begin{eqnarray}
\mathcal{F}_{C}^{\eta} &=&\sum_{i=1}^{n}\frac{1}{\mathcal{E}_{i}}\left( \frac{%
\partial \mathcal{E}_{i}}{\partial \eta }\right) ^{2},\text{ \ \ \ \ \ \ \ \
\ \ \ \ \ \ \ \ \ \ \ \ \ \ \ \ \ \ \ \ \ \ \ \ }\mathcal{E}_{i}\neq 0
\label{q3} \\
\mathcal{F}_{P}^{\eta} &=&4\sum_{i=1}^{n}\mathcal{E}_{i}\left( \left\langle \frac{%
\partial \mathcal{V}_{i}}{\partial \eta }\right\vert \left. \frac{\partial
\mathcal{V}_{i}}{\partial \eta }\right\rangle -
\expect{\mathcal{V}_{i}\Big|
\frac{\partial\mathcal{V}_{i}}{\partial
\eta }}\right), \nonumber\\
\mathcal{F}_{M}^{\eta} &=&8\sum_{i\neq j}^{n}\frac{\mathcal{E}_{i}\mathcal{E}_{j}}{%
\mathcal{E}_{i}+\mathcal{E}_{j}}
\Big|\expect{\mathcal{V}_{i}\Big|\frac{\partial
\mathcal{V}_{i}}{\partial \eta }}\Big|^2,
\text{ \ \ \ \ \ \ \ }\mathcal{E}_{i}+%
\mathcal{E}_{j}\neq 0  \notag
\end{eqnarray}

Using the eigenvalues $\epsilon_i$ and the eigenvectors
$\ket{V_i}$ of the density operator (4), one obtains  the explicit
form of Fisher information for arbitrary parameter $\eta$ as:

\begin{eqnarray}
\mathcal{F}_{C}^{\eta } &=&\sum_{i=1}^{4}\frac{1}{\epsilon_i}\left( \frac{%
\partial \epsilon_i}{\partial \eta }\right) ^{2},  \label{q4} \\
\mathcal{F}_{P}^{\eta } &=&\frac{4\epsilon_{3}\left[ \mu_{-}
^{\prime }{}^{2}+\mu_{-} ^{2}\theta ^{\prime 2}\right]
}{(1+\mu_{-} ^{2})^{2}}+\frac{4\epsilon_{4}\left[ \mu_{+} ^{\prime
}{}^{2}+\mu_{+} ^{2}\theta ^{\prime 2}\right] }{(1+\mu_{+}
^{2})^{2}},
\notag \\
\mathcal{F}_{M}^{\eta }
&=&\frac{8\epsilon_{3}\epsilon_{4}}{(\epsilon_{3}+\epsilon_{4})\left(
1+\mu_{+} ^{2}\right) ^{3}\left( 1+\mu_{-} ^{2}\right)
^{3}}\left\{ \left( \mu_{+} -\mu_{-} \right) ^{2}\left[ \left(
1+\mu_{-} ^{2}\right) ^{2}\mu_{+} ^{\prime 2}+\left( 1+\mu_{+}
^{2}\right) ^{2}\mu_{-} ^{\prime 2}\right] \right.  \notag \\
&&\left. +2\mu_{+} ^{2}\mu_{-} ^{2}\left( 1+\mu_{+} ^{2}\right) ^{2}\left( 1+\mu_{-}
^{2}\right) ^{2}\theta ^{\prime 2}\right\} ,
\end{eqnarray}
where $ \mu_{\pm} ^{\prime } =\frac{\partial \mu_{\pm} }{\partial \eta }$
 and $\theta
^{\prime }=\frac{\partial \theta }{\partial \eta }$. In the
following subsection, we investigate the initial parameters of the
charged qubit  $E_{j}$ and $E_{m}$ and the channel noisy parameter
$\Gamma$.

\subsection{Fisher information with respect to $\Gamma$}
In this subsection, we estimate the parameter $\Gamma$ by
evaluating the corresponding Fisher information, namely
$\mathcal{F}_{\Gamma}$. In Fig.(1a), we investigate the effect of
$E_i$ on the dynamics of $\mathcal{F}_{\Gamma}$. It is clear that,
at $t=0$, $\mathcal{F}_{\Gamma}$ is zero. However, as soon as the
interaction is switched on, the Fisher information increases
suddenly to reach its maximum value. For further values of $t$,
$\mathcal{F}_{\Gamma}$ decreases fast to reach its minimum value.
The minimum values depend on the values of $E_i$, where these
minimum values are larger for larger values of $E_i$. On the other
hand, this behavior is unchanged  for further values of $t$.

In Fig.(1b), we investigate the effect of different initial values
of $\Gamma$ on the $\mathcal{F}_{\Gamma}$, where we  set
$E_j=E_m=0.1$. It is evident that, for small values of $t$,  a
similar behavior is depicted as that shown in Fig.(1a), i.e., the
sudden increasing behavior is displayed as soon as the interaction
is switched on. Fisher information reaches its upper bound as $t$
increases. However, the upper bounds depend on the initial values
of $\Gamma$. It is clear that, as the initial values of $\Gamma$
increases, the  upper bounds of $\mathcal{F}_{\Gamma}$ increases.
For further values of $t$, the Fisher information decreases fast
to reach its lower bounds at different values of $t$ depending on
the initial values $\Gamma$. Also, $\mathcal{F}_{\Gamma}$ behaves
continently as the interaction time increases.

From Fig.(1), one concludes that, for small range of the
interaction time, one can estimate the parameter $\Gamma$, with
high probability. One can increase, the degree of estimation by
starting with a larger values of the parameter $\Gamma$ or $E_i$
parameters.  The larger  values of interaction time  has no effect
on the degree of estimation.

\begin{figure}
\begin{center}
\includegraphics[width=16pc,height=12pc]{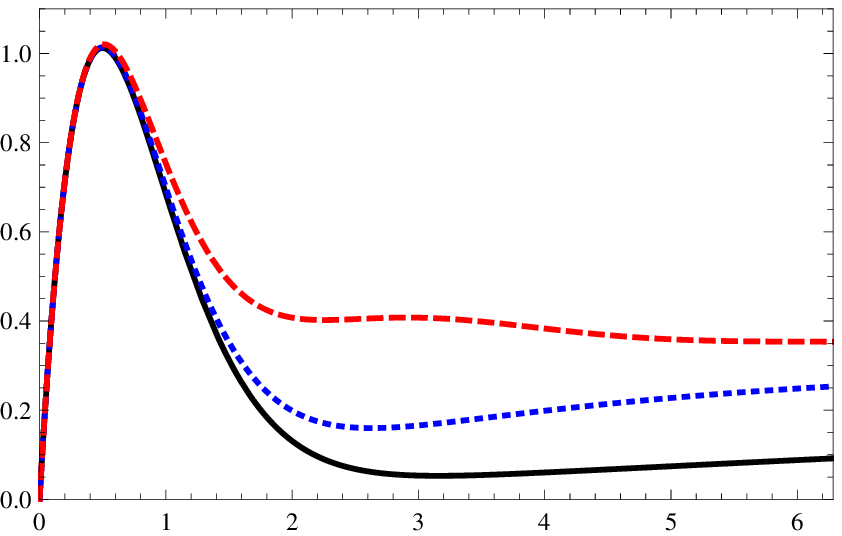}
\put(-210, 70){$\mathcal{F}_\Gamma$} \put(-25,120){$(a)$}
 \put(-100,-15){$t$}
~\quad\quad
\includegraphics[width=16pc,height=12pc]{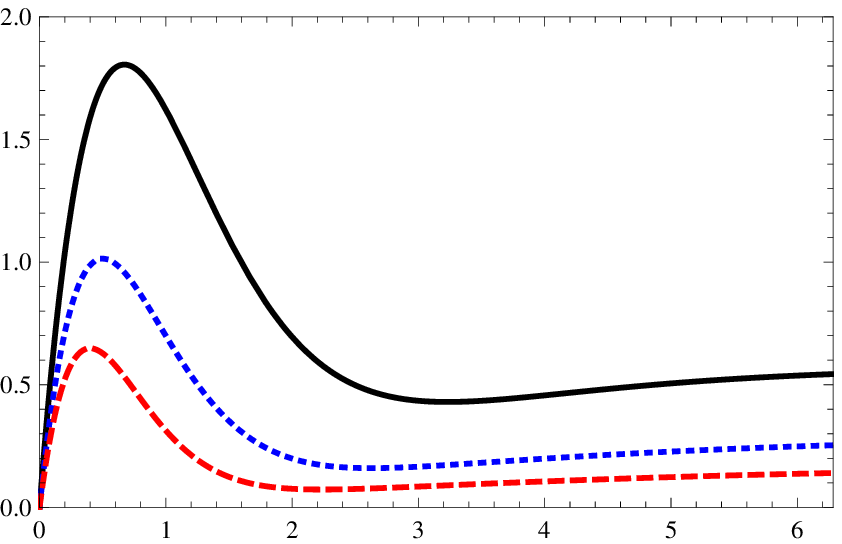} ~~~~~ %
 \put(-100,-15){$t$}
 \put(-45,120){$(b)$}
 \put(-230, 65){$\mathcal{F}_\Gamma$}
\end{center}
\caption{Fisher information $\mathcal{F}_{\Gamma }$ as a function
of time $t$.  In $(a)$, we set $\Gamma=0.4$ where, black-solid,
blue-dotted, red-dot dashed curves represent $E_{j}=E_{m}=0.05,
0.1, 0.2$, respectively. For $(b)$, we set $E_{j}=E_{m}=0.1$
where,   black-solid, blue-dotted, red-dotdashed curves represent
$\Gamma=0.3, 0.4, 0.5$, respectively. }
\end{figure}

In Fig.(2), we investigate the dynamics of the different parts of
the $\mathcal{F}_{\Gamma}$; the classical Fisher information
$\mathcal{F}_C$, the pure Fisher information $\mathcal{F}_P$ and
the mixture of pure state $\mathcal{F}_M$  on the behavior of
$\mathcal{F}_{\Gamma}$. It is clear that, the for small values of $t$ the
$\mathcal{F}_{C}$ has the the large contribution. However, as $t$
increases, the pure part $\mathcal{F}_P$ and the mixture part
$\mathcal{F}_M$ increase, therefor the Fisher information
$\mathcal{F}_{\Gamma}$ decreases.
 The effect of the energy parameters $E_j$, and the coupling
 parameter, $E_m$ can be seen by comparing Fig.(2a) and Fig.(2b), where the
the rate of increasing  $\mathcal{F}_P$ and $\mathcal{F}_M$ in
Fig.(2a) is smaller than that depicted in Fig.(2b). This shows
that as the energy and the coupling parameters increase, the
entangled charged state turns into a product pure states and the
possibility of construct the state of the charged qubit  as a
mixture of pure states increases.

From Fig.(2), we can see that as soon as the interaction is
switched on, the classical part $\mathcal{F}_C$ increases, namely,
the entangled qubits lose their quantum correlation. For larger
time there are some  pure states are generated and consequently
the pure Fisher information increases. Also, the initial state
behaves as a mixture of pure state, namely $\mathcal{F}_M$
increases.

\begin{figure}[t!]
\begin{center}
\includegraphics[width=16pc,height=12pc]{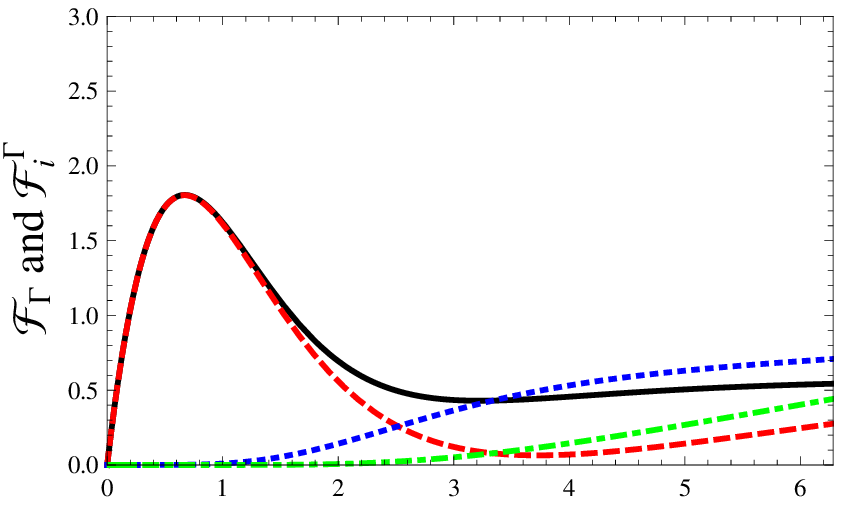}
 \put(-160,120){$(a)$}
 \put(-100,-15){$t$}
~\quad\quad
\includegraphics[width=16pc,height=12pc]{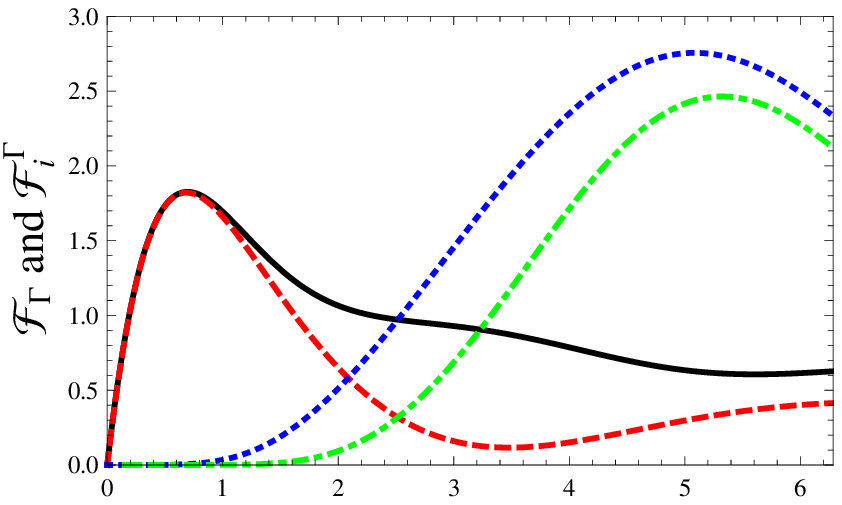}
\put(-100,-15){$t$}
 \put(-160,120){$(b)$}
\end{center}
\caption{ Fisher information $\mathcal{F}_{\Gamma }$ and its
components $\mathcal{F}^{\Gamma}_i$, $i=C,P,M$ as a function of
time $t$ when $\Gamma=0.4$. The black-solid, red-dotted,
blue-dashed and green-dotdashed curves for $\mathcal{F}_{\Gamma
}$,$\mathcal{F}_{C}^{\Gamma }$,$\mathcal{F}_{P}^{\Gamma }$ and
$\mathcal{F}_{M}^{\Gamma }$, respectively. We set in ${(a)}$
$E_{j}=E_{m}=0.1$ and in ${(b)}$ $E_{j}=E_{m}=0.2$. }
\end{figure}

\subsection{Fisher information with respect to $E_{j}$}
\begin{figure}
\begin{center}
\includegraphics[width=16pc,height=12pc]{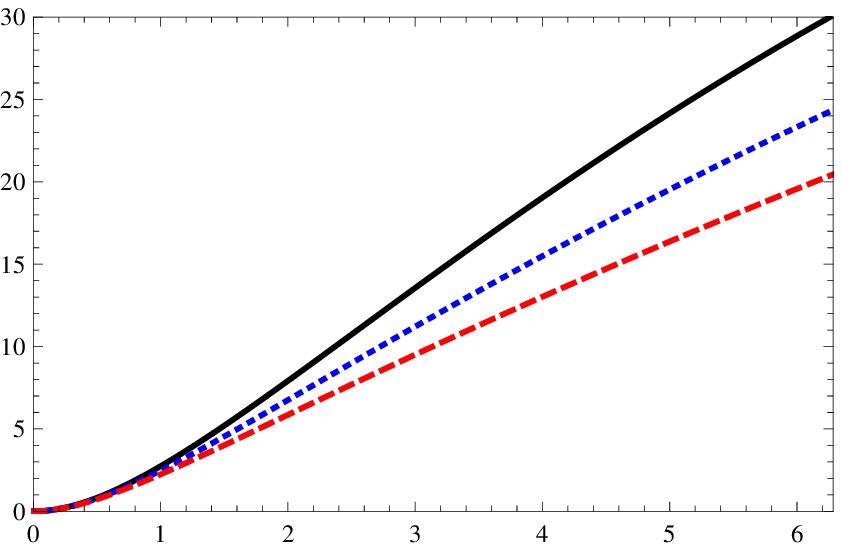}
\put(-210, 70){$\mathcal{F}_{E_{j}}$} \put(-170,120){$(a)$}
 \put(-100,-15){$t$}
~\quad\quad
\includegraphics[width=16pc,height=12pc]{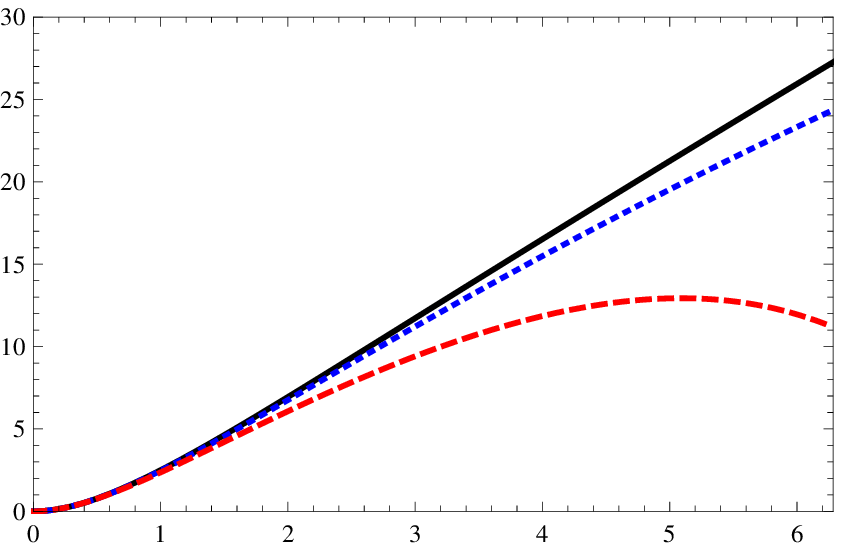}
\put(-210, 70){$\mathcal{F}_{E_{j}}$} \put(-170,120){$(b)$}
 \put(-100,-15){$t$}
\end{center}
\caption{Fisher information $\mathcal{F}_{E_{j}}$ as a function of
time $t$. The same parameters and properties in Fig. 1 are used }
\end{figure}

Fig.(3a) displays the behavior of  Fisher information
corresponding to the energy parameter $E_i$, where we set
$E_1=E_2=E$. In Fig.(3a), we fix the value of the depshing
parameter $\Gamma$ and  different initial energies are considered.
It is clear that, $\mathcal{F}_{E_{j}}$ increases as soon as the
interaction is switched on. The upper bounds depend on the initial
values of $E$, where for small initial values of $E$, the upper
bounds of Fisher information $\mathcal{F}_{E_{j}}$ are larger than those
depicted for small values of $E$.

The effect of different values of the depshing parameter $\Gamma$
is displayed in Fig.(3b), where different values of $\Gamma$ are
considered. It is clear that, as $\Gamma$ increases, the Fisher
information $\mathcal{F}_{E_{j}}$ decreases. Moreover, for small
values of the depashing parameter, Fisher information increases as
$t$ increases. However as $\Gamma$ increases further, the Fisher
information decreases.

The effect of the different part of the Fisher information are
displayed in Fig.(4a), where the classical part has the larger
contribution, while the pure and mixed parts have a small effect.
This behavior is changed dramatically as one increases the energy
parameter $E$, where   as  the interaction time $t$ increases, the
pure and the mixed parts of Fisher information  rapidly
increasing . This explain that why the upper bounds of the Fisher
information at $E=0,1$ is larger than that displayed at $E=0.2$.
Also, as it is shown in Fig.(4b), the charged qubit turns into
product pure state and the possibility of representing it as a
mixture of pure states increases as the interaction time
increases.

\begin{figure}
\begin{center}
\includegraphics[width=16pc,height=12pc]{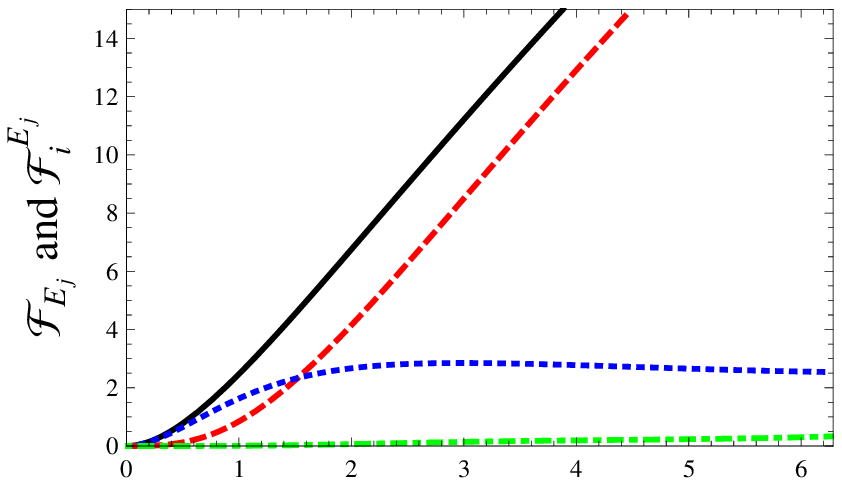}
\put(-160,120){$(a)$}
 \put(-100,-15){$t$}
~\quad\quad
\includegraphics[width=16pc,height=12pc]{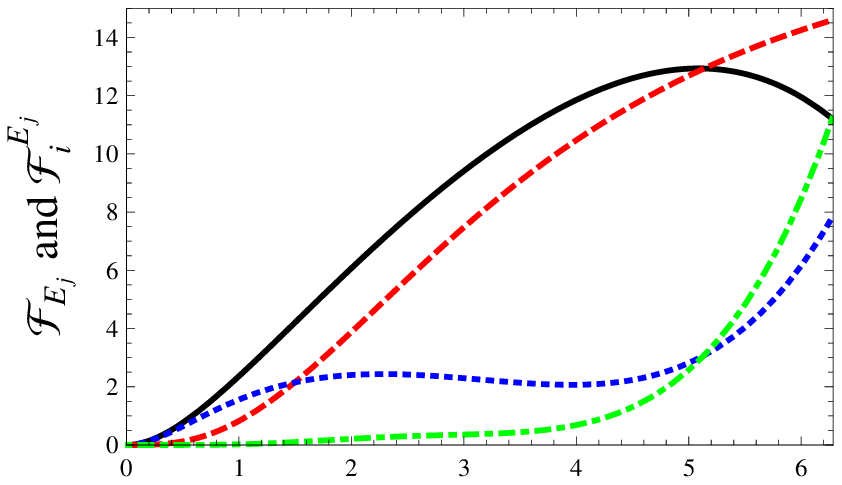}
 \put(-160,120){$(b)$}
 \put(-100,-15){$t$}
\end{center}
\caption{ Fisher information $\mathcal{F}_{E_{j}}$ and its components
$\mathcal{F}^{E_{j}}_i$, $i=C,P,M$  as a function of time $t$, with
the same parameters in Fig. 2 }
\end{figure}
\subsection{Estimation of the coupling parameter $E_{m}$}
\begin{figure}[t!]
\begin{center}
\includegraphics[width=16pc,height=12pc]{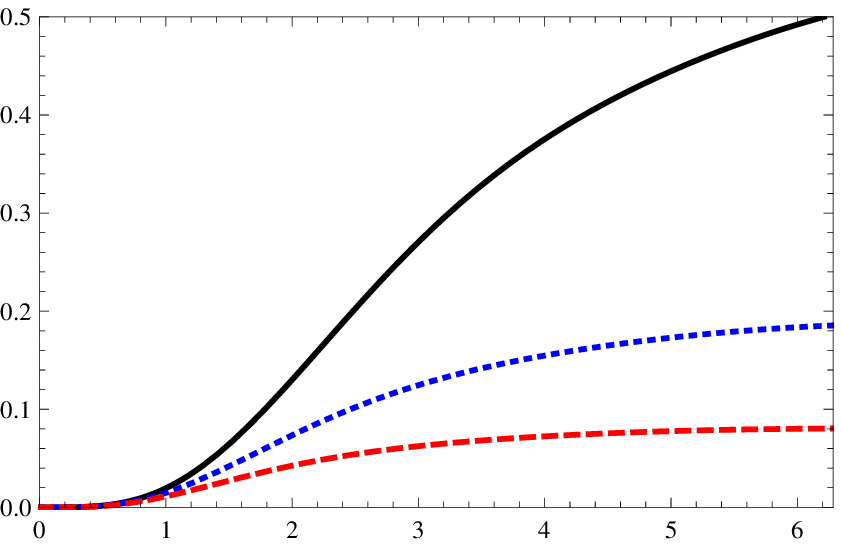}
\put(-210, 70){$\mathcal{F}_{E_m}$} \put(-170,120){$(a)$}
 \put(-100,-15){$t$}
~\quad\quad
\includegraphics[width=16pc,height=12pc]{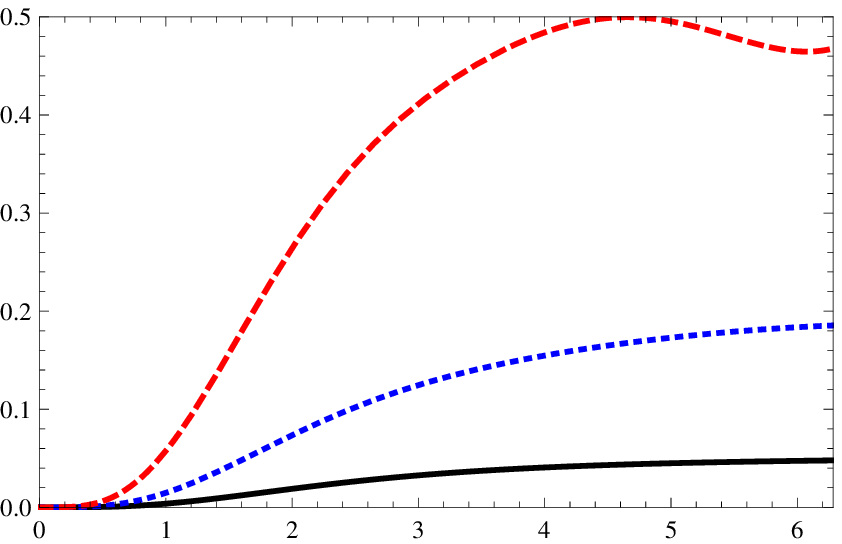}
 \put(-210, 70){$\mathcal{F}_{E_m}$} \put(-170,120){$(b)$}
 \put(-100,-15){$t$}
\end{center}
\caption{Fisher information $\mathcal{F}_{E_{m} }$ as a function
of time $t$. The same parameters and properties in Fig. 1 are used
}
\end{figure}

\begin{figure}
\begin{center}
\includegraphics[width=16pc,height=12pc]{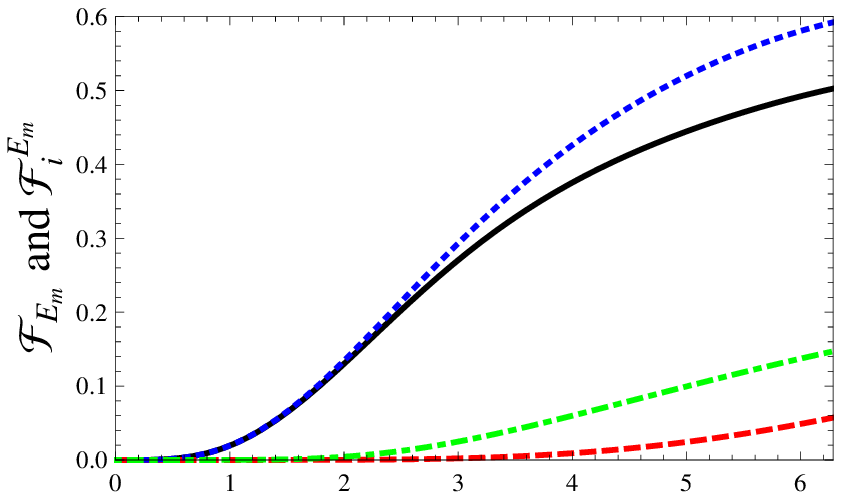}
\put(-160,120){$(a)$}
 \put(-100,-15){$t$}
~\quad\quad\quad
\includegraphics[width=16pc,height=12pc]{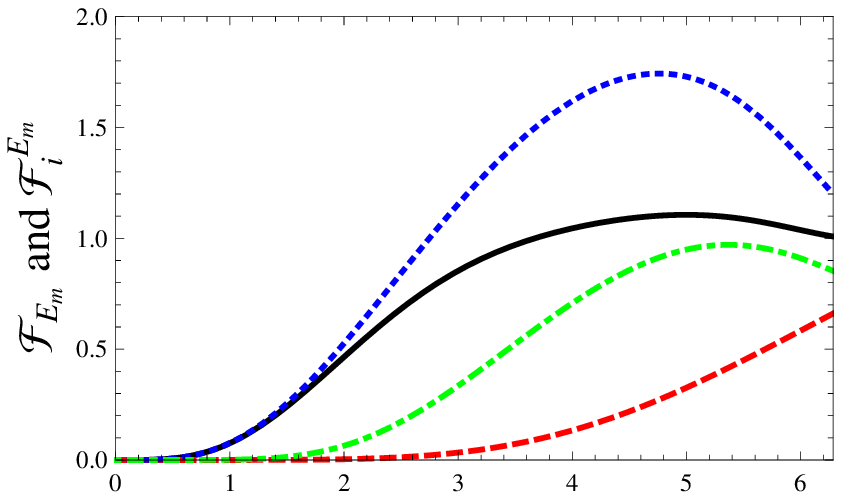}
\put(-160,120){$(b)$}
 \put(-100,-15){$t$}
\end{center}
\caption{ Fisher information $\mathcal{F}_{E_{m} }$ and its
components  $\mathcal{F}^{E_m}_i$, $i=C,P,M$  as a function of
time $t$, with the same parameters in Fig. 2 }
\end{figure}
To estimate the coupling parameter $E_m$, we evaluate the
corresponding Fisher information $\mathcal{F}_{E_m}$.  The effect
of different initial coupling on the Fisher information
$\mathcal{F}_{E_m}$ is displayed in Fig.(5a), where we fixed the
values of the dephasing end energy parameters. It is evident that,
as soon as the interaction is switched on, the Fisher information
increases $\mathcal{F}_{E_m}$ increases as $t$ increases. However,
the upper bounds of $\mathcal{F}_{E_m}$, depend on the coupling
parameter, where its upper bounds are small for larger values the
coupling parameter. For larger values of the energy parameter
$(E=0.2)$, the behavior is completely different. It is clear that,
the Fisher information increases as the initial values of the
coupling parameter increases (see Fig.(5b)). This shows that,  to
increase Fisher information of the coupling parameter, one has to
increase the  energy of the charged qubit and starting with a
larger value of the coupling parameter

The effect of the different parts of the Fisher information is
shown in Fig.(6a)  and Fig.(6b).  It is evident that, the pure
part $\mathcal{F}_P$ has the largest contribution, while the
classical part $\mathcal{F}_C$ has the smallest contribution.
However as one increases the energy and coupling parameters, the
classical part and the mixed part increases also for larger
interaction time, the pure part decreases. This explain the
decreases of the upper bounds of the Fisher information  in
Fig.(6b).

From Fig.(6) one can conclude that one can increase the
possibility of estimating the coupling  parameter  either by
starting with small values of this parameter with small energy, or
by larger initial value of the coupling and energy  parameters.
Also, the possibility of behaving the singlet state as pure
increases for smaller values of the energy and the coupling
parameters.

\section{conclusion}
In this contribution, we investigated the dynamics of a charged
two-qubit system  initially prepared in a maximum entangled state.
Each particle interacts independently with a noisy  dephasing
channel.  The final state of the charged qubits  was obtained
analytically. We estimated the channel parameter as well as the
energy parameters by means of Fisher information.

We showed that, Fisher information with respect to the energy
increases  suddenly as soon as the interaction is switched on to
reach its upper bounds. The upper bounds of Fisher information
depend on the interaction time, the initial values of the qubits'
energy  each the charged qubit. For further values of the
interaction time, the Fisher information decreases gradually,
where the decay rate increases as the energy increases.
The long-lived behavior  Fisher information with respect to the
noisy parameter is clearly depicted for larger values of
interaction time. Meanwhile, Fisher information with respect to
the energies of the charged qubits increases  suddenly as the
interaction time increases with larger values of the noisy
strength. On the other hand the behavior of Fisher information
with respect to the mutual coupling constant increases gradually
as the noisy strength increases.

The effect of the different parts of the Fisher  information is
discussed for all cases. We showed that classical part has the
larger contribution on the total Fisher information with respect
to the nosy and energy parameters, while for the mutual coupling
the pure part has the largest effect. This result is changed for
larger values of the initial energy, where the classical has a
largest effect on the mutual parameter's information.

From the behavior of the differen parts of Fisher information, one
deduced when the final state turns into a separable pure state. It
is clear that, the initial charged qubit system is robust against
the larger energy, since there still a contribution from the pure
and mixed parts.

\end{document}